\documentclass[a4paper,twosides]{article}
\usepackage{CJK,multicol,multirow,graphics,fancyhdr,epstopdf}
\usepackage{amsmath,amsfonts,amssymb,bm,upgreek,mathrsfs,ccmap,mathcomp}
\usepackage[pagewise,switch,columnwise]{lineno}
\usepackage[compress,nospace]{cite}
\usepackage[dvipsnames]{xcolor}
\usepackage{CPL-2023}

\newcommand{\cplyear}{2025} \newcommand{\cplvol}{xx}

\newcommand{\cplno}{x} \newcommand{\cplpagenumber}{xxxxxx}
 \newcommand{\cplpage}{\cplpagenumber-\thepage}

\begin{document}

\begin{CJK}{GBK}{song}\vspace* {-4mm} \begin{center}
\large\bf{\boldmath{Modeling of van-der-Meer scan at NICA}}
\footnotetext{\hspace*{-5.4mm}$^{*}$Corresponding authors. Email: krass58ad@mail.ru

}

\normalsize \rm{}Anton Babaev$^{1}$
\\[3mm]\small\sl $^{1}$Tomsk Polytechnic University, Tomsk 634050, Russia

\end{center}
\end{CJK}
\vskip 1.5mm

\small{\narrower
Van-der-Meer method is used in hadron colliders for absolute luminosity calibration. In the letter the applicability of the method is discussed when particles desnsity is distorted by hour-glass effect next to interaction point. The study is motivated by close commisioning of Nuclotron based Ion Collider fAcility where hour-glass effected is significant by design. The theoretical van-der-Meer formalism is revised for this case. Importance of hour-glass effect for luminosity measurements and calibration is demonstrated. Also, the general approach to the estimation of van-der-Meer method bias when it is used for non-factorizable beams is proposed. 
\par}


\vskip 5mm

The luminosity in high energy physics experiments is the measure of performance of collider. It characterizes the total number of collisions and amount of data collected for analyses.
For the purpose of this letter it is enough to be limited by the most common case of equal energy relativistic beams. The crossing angle isn't considered in this letter. With these preliminary remarks the luminosity at collision of two opposite bunches is determined by the overlap integral \cite{c2}:
\begin{equation}\label{eq3}
	L=2\hat{\beta}cN_1N_2\int\varrho_1(x,y,s,t)\varrho_2(x,y.s.t)dxdydsdt
\end{equation} 
where $\hat{\beta}$ is common longituginal velocity of particles in a bunch in units of speed of light $c$, $N_{1,2}$ are numbers of particles in bunches and $\varrho_{1,2}$ are densities of particles in bunches.
 Particles density depends on transverse coordinates $x$, $y$, longitudinal coordinate $s$, and integration is performed over time of interaction $t$  (space distribution is normalized by 1 at any time moment: $\int\varrho(x,y,s,t)dxdyds=1$). The definition \eqref{eq3} demonstrates that the luminosity, in general, is proportional to the probability of particles collision. Colliders deliver beams of certain properties to interaction point (IP) and then products of collision are registered by detector (see in recent review \cite{c4} for details). If $N_{ev}$ is the number of  reconstructed events corresponding some process of particle physics with cross-section $\sigma$ then $N_{ev}=\sigma{}L_{int}$ where $L_{int}$ is integrated luminosity collected for the dataset used for analysis.
 For curcular colliders which deliver bunched beams with the revolution frequency $f$ the intergated luminosity is $L_{int}= \int{}L_{inst}dt$  where $L_{inst}= fn_bL$ is instanteneous luminosity, $n_b$ is the number of colliding bunch pairs and all bunches in a beam are assumed identical.

The beam overlap can not be measured directly. Usually for fast online measurements the dedicated detector (luminometer) is constructed \cite{c5,c6,c7}. A luminometer registrates flux of secondary particles. Its integrated response per one bunch crossing is noted here $R$. To translate the detector's response $R$ to real luminosity $L$ the proper calibration is needed. It is convinient to consider the response is linear and the calibration constant $\sigma_{vis}$, so-called visible cross-section, is introduced: $R=\sigma_{vis}L$ \cite{c8}. 
For hadron colliders there is the experimental method to measure $\sigma_{vis}$, van-der-Meer (vdM) scan \cite{c8,c9}, is actively used at Large Hadron Collider (LHC) \cite{c10,c11,c31} and The Relativistic Heavy Ion Collider (RHIC) \cite{c12}. The using of this method is also proposed for Nuclotron based Ion Collider fAcility (NICA) \cite{c7}. NICA is expected to be commissioned in 2026 in Joint Institute for Nuclear Research (Dubna, Russia).

It is expected, in full operation NICA will deliver beams of various nuclei up to Au with center-of-mass collision energies $\sqrt{s_{NN}}$ from 4 to 11 GeV per nucleon \cite{c1} to Multi Purpose Detector (MPD) experiment for studies of QCD matter under extreme densities and temperatures \cite{c32,c21}. It is expected for Au+Au collisions at maximum energy NICA/MPD will get the instanteneous luminosity $\sim1E27$ cm$^{-2}$c$^{-1}$ \cite{c22} that is comparable to the luminosity at nuclei collision at LHC Compact Muon Solenoid experiment \cite{c23}.

The benefit of van-der-Meer method is for it requires only measurements of luminometer and information on beam conditions. The reconstruction of events which generate the response $R$ isn't required. In van-der-Meer method the response $R$ is measured in dependence on the separation between beam orbits (beam separation) during two orthogonal scans in transverse collider plane XY. 
For the precision of the experiment the assumption about  factorization of transverse distribution $\rho(x,y)$ of particles in a bunch is important, i.e. $\rho(x,y)=\rho(x)\rho(y)$ where $\rho(x,y)=\int\varrho{}ds$ (at fixed time moment). The transverse particle motion always more or less coupled in real beams, and the understanding of beam transverse non-factorization and estimation of corresponding uncertainty is important challenge in van-der-Meer analysis \cite{c10}. Following theoretical formalism of van-der-Meer method, the first quantity determined from the scan is the effective scan width\footnote{There is no conventional term for this quantity. For example, it can be called convolved beam size or effective width of beam overlap \cite{c31,c15} assuming head-on collisions, but it isn't valid for arbitrary beams. Sometime, as for q-Gaussian factorizable beams \cite{c16}, effective scan width can be adjusted to convolved beam size by proper tuning of constant factors in \eqref{eq1}, \eqref{eq6}. In this letter standard formulae are used but not the terms referring to bunch size because they might be misleading here.}
\begin{equation}\label{eq1}
\Sigma_x=\frac{1}{\sqrt{2\pi}}\frac{\int{}R(\Delta_x,0)d\Delta_x}{R(0,0)},\quad\Sigma_y=\frac{1}{\sqrt{2\pi}}\frac{\int{}R(0,\Delta_y)d\Delta_y}{R(0,0)}
\end{equation}
where $\Delta_{x,y}$ is beam separation in $x,y$ direction correspondingly and two scans are considered: X-scan at $\Delta_y=0$ and Y-scan at $\Delta_x=0$. The effective scan width doesn't depend directly on certain bunch density but only on scan curve shape. For transversely factorizable bunches with Gaussian longitudinal distribution it could be shown that the luminosity at head-on collisions:
\[
	L(0,0)=\frac{N_1N_2}{2\pi\Sigma_x\Sigma_x}
\]
and, therefore, visible cross-section, measured in van-der-Meer scan:
\begin{equation}\label{eq6}
	\sigma_{vis,vdm}=2\pi\frac{R(0,0)}{N_1N_2}\Sigma_x\Sigma_y
\end{equation}

The transverse beam size $\sigma_{x,y}$ in collider is determined by the lattice and can be expressed using transverse beam emittance $\varepsilon_{x,y}$ and envelop of betatron oscillations $\beta_{x,y}$ ($\beta$-function): $\sigma_{x,y}=\sqrt{\varepsilon_{x,y}\beta_{x,y}}$ \cite{c2,c8}. In sections of collider where $\beta$-function is constant the particle distribution is usually well described by Gaussian distribution. However, at the interaction point (IP) for higher luminosity the transverse beam size should be made as small as technically possible. So, next to IP $\beta$-function has the parabolic shape \cite{c8}:
\[
	\beta_{x,y}(s)=\beta_{x,y}^{*}\left(1+\frac{s^2}{(\beta_{x,y}^{*})^2}\right)
\] 
where $s$ is the distance from the IP and $\beta^{*}$ is the value of $\beta$-function at IP. Below, as for most practical cases, $\beta_x^{*}=\beta_y^{*}=\beta^{*}$. Therefore, the beam size $\sigma_{x,y}$ next to IP is changed and, for long bunches ($\alpha=\sigma_s/\beta^{*}\geq1$, $\sigma_s$ is the bunch length), the beam size can not be considered constant while bunches interact. This effect at head-on collision reduces luminosity and it is known as hour-glass (HG) effect \cite{c2,c33,c19}:
\[
	L(0,0)=\frac{N_1N_2}{4\pi\sigma^{*}_x\sigma^{*}_y}\Phi_{HG}(\alpha)
\]
where
\[
	\Phi_{HG}(\alpha)=\frac{2}{\sqrt{\pi}}\int_0^{\infty}\frac{e^{-u^2}}{1+(\alpha{}u)^2}du
\]
is the hour-glass factor, $\sigma^{*}_{x,y}$ is transverse beam size at IP. The density of Gaussian bunch distorted by HG effect can be written in the form \cite{c19}
\begin{equation}\label{eq2}
	\rho(x,y,\eta,s_0)=\frac{1}{(2\pi)^{3/2}\sigma_x(s_0,\eta)\sigma_y(s_0,\eta)\sigma_s}\exp\left[-\frac{1}{2}\left(\frac{x^2}{(\sigma_x(s_0,\eta))^2}+\frac{y^2}{(\sigma_y(s_0,\eta))^2}+\frac{\eta^2}{\sigma_s^2}\right)\right]
\end{equation}
where $s_0$ is the longitudinal coordinate of bunch center, $\eta$ is the longitudinal coordinate of a particle in the bunch relative to the center, $s=s_0+\eta$, $\rho(x,y,\eta,s_0)=\varrho(x,y,s_0+\eta,s_0/(\hat{\beta}{}c))$. Transverse bunch size can be written in the form:
\begin{equation}\label{eq7}
	\sigma_{x,y}(s_0,\eta)=\sigma_{x,y}^{*}\sqrt{1+\frac{(s_0+\eta)^2}{\beta^{*2}}}.
\end{equation}

The HG effect is important at NICA where $\alpha=1$ (see in Table~1) and, therefore, $\Phi_{HG}\approx0.76$. For illustration, in Fig.\,1 $(x,s)$ distribution $\rho(x,\eta,s_0)=\int\rho(x,y,\eta,s_0)dy$ is shown for three time moments: (a) before the bunch center passes IP, (b) at the moment when it passes IP, and (c) after it passes IP. The bunch gets tails with the width evolving while the bunch passes the interaction region and crosses the similar opposite bunch. This is the cause of HG effect in luminosity.

\vskip 4mm

\fl{1}\centerline{\includegraphics{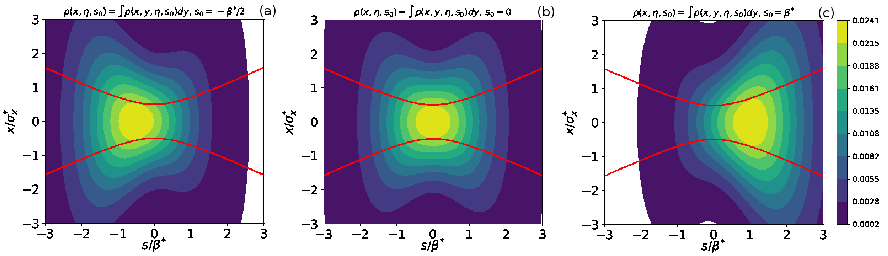}}

\vskip 2mm

\figcaption{15.5}{1}{Evolution of $(x,s)$ distribution while the bunch \eqref{eq2} passes interaction region from (a) to (c): (a) $s_0=-\beta^{*}/2$, (b) $s_0=0$, (c) $s_0=\beta^{*}$. Colour scale represents $\rho(x,\eta,s_0)=\int\rho(x,y,\eta,s_0)dy$ (in cm$^{-2}$ units). Red lines represents $\pm\sigma_x(s)/2$ curves \eqref{eq7}. Plots are for NICA conditions, see in Table~1.}

\medskip

\newpage

\vskip 2mm

\tl{1}\tabtitle{7.8}{1}{Parameters of NICA collider used in the letter for Au collisions at $\sqrt{s_{NN}}=11$ GeV, adopted from \cite{c7}.}

\vskip 2mm \tabcolsep 4.5pt

\centerline{\footnotesize
\begin{tabular}{cc}
\hline\hline\hline
$\hat{\beta}$ & 0.985 \\
 $\sigma_s$, $\beta^{*}$, cm & 60 \\ 
$f$, c$^{-1}$ & 5.87E5 \\
$n_b$ & 22 \\
$N_1=N_2$ & 2.8E9 \\
$\sigma^{*}_x$, mm & 1.1 \\
$\sigma^{*}_y$, mm & 0.82 \\
\hline\hline\hline
\end{tabular}}

\vskip 2mm

\medskip

Hence, concerning colliders like NICA the question arises on how HG effect will influence luminosity not only at head-on collision but in vdM scan in general, when beams are separated at the distance up to few $\sigma_{x,y}^{*}$, and how it changes quantities measured in vdM scan ($\Sigma_{x,y}$, $\sigma_{vis,vdm}$). The luminosity at overlap of two bunches \eqref{eq2} of equal dimensions $(\sigma^{*}_x,\sigma^{*}_y,\sigma_s)$ at the beam separation $(\Delta_x,\Delta_y)$ can be written following equation \eqref{eq3}:
\[
	L(\Delta_x,\Delta_y)=2N_1N_2\int\rho_1(x,y,\eta,s_0)\rho_2(x+\Delta_x,y+\Delta_y,\eta+2s_0,-s_0)dxdyd\eta{}ds_0
\] 
where integration over longitudinal coordinate and interaction time is replaced by integration over $\eta$ and $s_0$ for first bunch, and centers of both bunches pass IP simultaneously. With the procedure similar to \cite{c19} one can obtain
\begin{equation}\label{eq4}
	L(\Delta_x,\Delta_y)=L_{Gauss}(\Delta_x,\Delta_y)\Phi_{HG,vdm}(\Delta_x,\Delta_y,\alpha).
\end{equation}
So, luminosity as a function of beam separation in colliders with HG effect is decomposed into the product of luminosity for Gaussian bunches:
\begin{equation}\label{eq5}
	L_{Gauss}(\Delta_x,\Delta_y)=\frac{N_1N_2}{4\pi\sigma_x^{*}\sigma_y^{*}}\exp\left(-F(\Delta_x/\sigma_x^{*},\Delta_y/\sigma_y^{*})\right)
\end{equation}
and the factor which describs dependence of HG factor on beam separation
\begin{equation}\label{eq8}
	\Phi_{HG,vdm}(\Delta_x,\Delta_y,\alpha)=\frac{2}{\sqrt{\pi}}\int_0^{\infty}\exp\left[-u^2\left(1-\frac{\alpha^2F(\Delta_x/\sigma_x^{*},\Delta_y/\sigma_y^{*})}{1+(\alpha{}u)^2}\right)\right]\frac{du}{1+(\alpha{}u)^2}
\end{equation}
where
\[
	F(\xi,\chi)=\frac{1}{4}\left(\xi^2+\chi^2\right).
\]
In particular, $\Phi_{HG,vdm}(0,0,\alpha)=\Phi_{HG}(\alpha)$ is the ususal HG factor, and, if $\alpha=0$ then $L(\Delta_x,\Delta_y)=L_{Gauss}(\Delta_x,\Delta_y)$. In Fig.\,2(a) the luminosity \eqref{eq4} multiplied by $fn_b$ is plotted for expected NICA conditions, see in Table~1. 
For comparasion the similar plot for Gaussian beams \eqref{eq5} with the same intensity and dimensions is provided. In experiment there will be similar curves for response of luminometer, $R\propto{}L$. The HG effect in NICA significantly distorts the scan curve: it reduces expected luminosity at small beam separation, but at large separations the HG factor enhances luminosity. In Fig.\,2(b) the HG factor $\Phi_{HG,vdm}(\Delta_x,0,\alpha)$ is shown for few $\alpha$ parameters. For NICA conditions the factor in changed three times within the range of interest.

\vskip 4mm

\fl{2}\centerline{\includegraphics{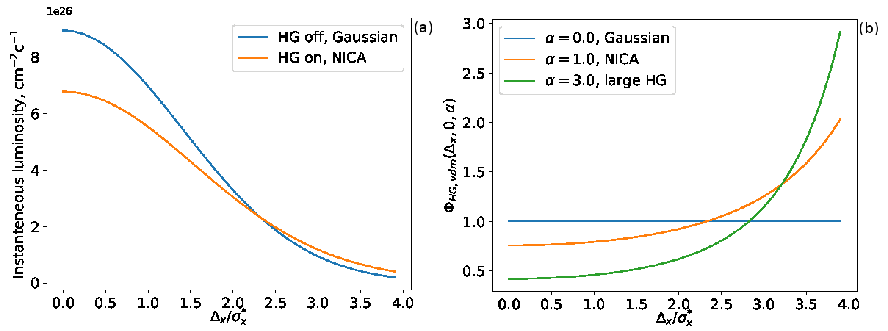}}

\vskip 2mm

\figcaption{14.0}{2}{(a) Instanteneous luminosity in X vdM scan at NICA conditions and for Gaussian beams of the same $N_{1,2}$, $\sigma^{*}_{x,y}$. (b) HG factor $\Phi_{HG,vdM}$ \eqref{eq8} in X vdM scan for Gaussian beams (constant), at NICA conditions and at large HG effect.}

\medskip

Using equations \eqref{eq4} and \eqref{eq1} one can find equations for effective scan width:
\begin{equation}\label{eq9}
	\Sigma_{x,y}=\frac{2\sigma_{x,y}^{*}}{\sqrt{2\pi}\alpha\Phi_{HG}(\alpha)}\exp\left(\frac{1}{2\alpha^2}\right)K_0\left(\frac{1}{2\alpha^2}\right)
\end{equation}    
where $K_0(\xi)$ is the modified Bessel function of 0-th order. For NICA this equation gives $\Sigma_{x,y}\approx1.6045\sigma_{x,y}^{*}$ ($\Sigma_x\approx0.18$ cm, $\Sigma_y\approx0.13$ cm). The corresponding values for Gaussian beams of the same size at IP are $2^{1/2}\sigma_{x,y}^{*}$ \cite{c8}. It means that HG effect increases scan curve width at $\approx13\%$ that confirms importance of HG effect in vdM analysis and for precise luminosity calibration. 

Equation~\eqref{eq9} connects beam size at IP and scan curve width. In experiment the scan curve width is often determined directly using analytical functions to fit the measured $R$ where $\Sigma_{x,y}$ are fit parameters \cite{c10,c12,c16}. Equation \eqref{eq4} shows that this approach isn't well suited to take into account the HG-related distortion of scan curve shape. The problem of best fit function is out of scope of this letter, the similar task was considered recently in \cite{c16} for q-Gaussian beams.


Above it was noted the vdM method gives unbiased results only when beams are factorizable in transverse plane. For bunches with distribution \eqref{eq2} this condition isn't valid. The bias $r_{nf}$ can be estimated from the ratio of experimentally measured $\sigma_{vis,vdm}$ \eqref{eq6} and true visible cross-section given following its definition $\sigma_{vis}=R(0,0)/L(0,0)$ at the same response of luminometer: $r_{nf}=\sigma_{vis,vdm}/\sigma_{vis}$.  So, for colliders with HG effect this bias is
\[
	r_{nf}(\alpha)=\frac{e^{1/\alpha^2}}{\pi\alpha^2\Phi_{HG}(\alpha)}K_0^2\left(\frac{1}{2\alpha^2}\right)
\]
where $L(0,0)$ is given by \eqref{eq4}. For NICA conditions $r_{nf}(1)\approx0.9756$, so $\sigma_{vis,vdm}$ is underestimated at $\approx2\%$. Therefore, the luminosity calibrated with using the visible cross-section determined by vdM method will be overestimated at $\approx2\%$. The corresponding correction can be be introduced directly at operation. It should be noted the proposed method for estimation of non-factorisation bias can be implemented for any kind of beams, not only for distorted by HG effect. Another important remark is that this bias originates from the density model; when collider will be in operations there could be another issues cause beam non-factorisation, for example, beam-beam interactions \cite{c15}. 

To summarize, in this letter the vdM formalism for colliders with significant HG effect is considered. The interest to this work rises due to close commissioning of NICA collider where vdM metod is proposed for luminosity calibration \cite{c7}. 
The theoretical formalism that vdM method is based on is revised. As a result it is shown HG effect significantly varies within range of beam separations and it noticeable influences vdM analysis. It should be stressed HG effect isn't important for experiments based at LHC because LHC delivers very short bunches ($\sigma_s<10$ cm) whereas accelerator settings provide $\beta^{*}>10$ m especially for vdM scan in proton-proton program \cite{c10} and $\beta^{*}>30$ cm in nuclei-nuclei runs \cite{c23}. On the contrary, at RHIC the head-on HG factor is comparable with considered in this letter, the change of HG factor with beam separation was considered in (RHIC papers use the term ``Vernier scan'' instead of vdM scan) with, in general, the same result \cite{c12}.

Another important topic of the letter is the non-factorization problem. The general scheme for bias estimation is proposed and estimations for NICA also presented.

In this letter only HG effect and its appearance in vdM method are considered. Many operational issues in luminosity measurements and calibration are not mentioned, they are well described in \cite{c4,c10,c15}. These issues also combine the HG effect and, probably, more detailed studies will be needed when NICA will be in operations.

\textit{Acknowledgements.} This work was supported by The Ministry of Science and Higher Education of the Russian Federation in part of the Science program (Project No. FSWW-2023-0003)

\end{document}